\documentclass[letter,twocolumn]{jpsj3}
\usepackage{txfonts}
\usepackage{graphicx}
\usepackage{dcolumn}
\usepackage{bm}

\def\lesssim{\ \raise.3ex\hbox{$<$}\kern-0.8em\lower.7ex\hbox{$\sim$}\ }
\def\gesim{\ \raise.3ex\hbox{$>$}\kern-0.8em\lower.7ex\hbox{$\sim$}\ }

\title{Dip-hump temperature dependence of Specific Heat and Effects of Pairing Fluctuations in the Weak-coupling Side of a $p$-wave Interacting Fermi Gas}

\author{Daisuke Inotani\thanks{dinotani@rk.phys.keio.ac.jp}, Pieter van Wyk, Yoji Ohashi}
\inst{Department of Physics, Keio University, 3-14-1 Hiyoshi, Kohoku-ku, Yokohama 223-8522, Japan} 

\abst{We investigate the specific heat $C_V$ at constant volume in the normal state of a $p$-wave interacting Fermi gas. Including $p$-wave pairing fluctuations within the strong-coupling theory developed by Nozi\`eres and Schmitt-Rink, we show that, in the weak-coupling side, $C_V$ exhibits a dip-hump behavior as a function of the temperature. While the dip is associated with the pseudogap phenomenon near $T_{\rm c}$, the hump structure is found to come from the suppression of Fermi quasiparticle scattering into a $p$-wave molecular state in the Fermi degenerate regime. Since the latter phenomenon does not occur in the ordinary $s$-wave interacting Fermi gas, it may be viewed as a characteristic phenomenon associated with a $p$-wave pairing interaction.
} 

\begin{document}
\maketitle
\par
In an ultracold Fermi gas with a $p$-wave pairing interaction, the most characteristic phenomenon is the $p$-wave superfluid phase transition. Since the realization of a tunable $p$-wave interaction associated with a Feshbach resonance, extensive experimental\cite{Regal,Regal2,Ticknor,Zhang,Schunck,Gunter,Gaebler2,Inaba,Fuchs,Mukaiyama,Maier} and theoretical efforts\cite{Gurarie,Ohashi,Ho,Botelho,Iskin,Iskin2,Cheng,Levinsen,Gurarie2,Grosfeld,Iskin3,Mizushima,Han,Mizushima2,Inotani,Inotani2} have been done to realize this unconventional Fermi superfluid. However, although the formation of $p$-wave molecules has been reported\cite{Gaebler2}, no one has succeeded in this exciting challenge. This is because, although a $p$-wave interaction is always needed to realize a $p$-wave superfluid Fermi gas, it also causes the dipolar relaxation\cite{Regal2,Gaebler2,Inaba,Fuchs}, as well as three-body loss\cite{Castin,Gurarie3}, leading to very short lifetime ($\tau_{\rm Cooper}=5\sim 20$ ms) of $p$-wave Cooper pairs\cite{Chevy}, as well as particle loss from the system. As a result, these pairs are soon destroyed before the condensation growth ($=O(100~{\rm ms})\gg \tau_{\rm Cooper}$). 
\par
Because of this difficulty, in the current stage of research on a $p$-wave interacting Fermi gas, it is interesting to look for other characteristic phenomena that are absent in the ordinary $s$-wave case. Since the current experiment is only accessible to the normal state of a $p$-wave interacting Fermi gas, one should explore such phenomena above $T_{\rm c}$. To avoid the serious particle loss\cite{Regal2,Gaebler2,Inaba,Fuchs,Castin,Gurarie3} as much as possible, the weak-coupling regime away from a $p$-wave Feshbach resonance would be better.  
\par
In this letter, we present a phenomenon that meets the above demands, within the framework of the strong-coupling theory developed by Nozi\`eres and Schmitt-Rink. Usually, interaction effects are considered to be weak in the weak-coupling regime. However, even in this regime, we show that a $p$-wave pairing interaction still gives an anomalous temperature dependence of the specific heat $C_V$ at constant volume above $T_{\rm c}$, which is not seen in the $s$-wave case at all. We briefly note that the observation of this thermodynamic quantity has recently become possible in cold Fermi gas physics\cite{Ku}.
\par
We consider a one-component uniform Fermi gas with a $p$-wave interaction, described by the Hamiltonian\cite{Ohashi,Ho,Inotani,Inotani2},
\begin{equation}
H=\sum_{\bm p} \xi_{\bm p}c_{\bm p}^{\dagger}c_{\bm p}
-\frac{1}{2}\sum_{{\bm p},{\bm p}',{\bm q}} 
V_p({\bm p},{\bm p}')
c_{{\bm p}+{\bm q}/2}^\dagger c_{-{\bm p}+{\bm q}/2}^\dagger
c_{-{\bm p}'+{\bm q}/2}c_{{\bm p}'+{\bm q}/2}.
\label{eq.1}
\end{equation}
In this letter, we set $\hbar=k_{\rm B}=1$, and the system volume $V$ is taken to be unity, for simplicity. Here, $c_{\bm p}$ is an annihilation operator of a Fermi atom with the kinetic energy $\xi_{\bm p}=p^2/(2m)-\mu$, measured from the Fermi chemical potential $\mu$ (where $m$ is an atomic mass). In Eq. (\ref{eq.1}), 
\begin{equation}
V_p({\bm p},{\bm p}')=-\sum_{i=x,y,z} \gamma^i_{\bm p} U \gamma^i_{{\bm p}'}
\label{eq.1b}
\end{equation}
is an assumed $p$-wave pairing interaction\cite{Ohashi,Ho,Inotani,Inotani2}, where$U~(>0)$ is a coupling constant. The $p$-wave basis functions $\gamma^i_{\bm p}=p_i F_{\rm c}({\bm p})$ ($i=x,y,z$) involve a cutoff function $F_{\rm c}({\bm p})$, which will be specified soon later. In a real ultracold Fermi gas, a $p$-wave interaction can be tuned by adjusting the threshold energy of a Feshbach resonance\cite{Chin}. However, we do not consider the detailed Feshbach mechanism in this letter, but simply treat $U$ as a tunable parameter. As observed in a $^{40}$K Fermi gas\cite{Ticknor}, a $p$-wave interaction in an ultracold Fermi gas may have a uniaxial anisotropy ($U_x \ge U_y= U_z$, where the $x$ axis is chosen to be parallel to an external magnetic field to adjust a Feshbach resonance), because of the splitting of a $p$-wave Feshbach resonance by a magnetic dipole-dipole interaction. However, we ignore this, to only deal with the isotropic case ($U_x=U_y=U_z\equiv U$), for simplicity. Effects of the uniaxial anisotropy will be separately discussed in our future paper. We also ignore effects of a harmonic trap. 
\par
The cutoff function $F_{\rm c}({\bm p})$ in the basis function $\gamma^i_{\bm p}$ is to eliminate the well-known ultraviolet divergence involved in the model Hamiltonian in Eq. (\ref{eq.1}). Here, we take $F_{\rm c}({\bm p})=1/[1+(p/p_{\rm c})^{2n}]$ with $n=3$\cite{Inotani2}. We briefly note that, while superfluid properties below $T_{\rm c}$ somehow depend on the value of $n$, normal state properties above $T_{\rm c}$ do not\cite{Inotani2}, as far as the cutoff momentum $p_0$ is taken to be much larger than the Fermi momentum $k_{\rm F}$. Although the momentum cutoff $p_{\rm c}$ and the bare $p$-wave interaction strength $U$ are not observable, they are related to the observable scattering volume $v$, as well as the inverse effective range $k_0$, as
\begin{equation}
{4\pi v \over m}=-
{U \over 3}
{1 \over \displaystyle 1-{U \over 3}\sum_{\bm p}{p^2 \over 2\varepsilon_{\bm p}}F_{\rm c}^2({\bm p})},
\label{eq.3}
\end{equation}
\begin{equation}
k_0=-{4\pi \over m^2}
\sum_{\bm p}{p^2 \over 2\varepsilon_{\bm p}^2}F_{\rm c}^2({\bm p}).
\label{eq.4}
\end{equation}
We take $k_0=-30k_{\rm F}$ (where $k_{\rm F}$ is the Fermi momentum), following the experiment on a $^{40}$K Fermi gas\cite{Ticknor}. When we measure the strength of a $p$-wave interaction in terms of the inverse scattering volume $v^{-1}$, the weak-coupling side and the strong-coupling side are conveniently characterized as $(k_{\rm F}^3v)^{-1}\lesssim 0$ and $(k_{\rm F}^3v)^{-1}\gesim 0$, respectively. We briefly note that, although there is actually no clear boundary between the two regions, the Fermi chemical potential $\mu(T\simeq T_{\rm c})$ becomes negative around $(k_{\rm F}^3v)^{-1}= 0$\cite{Ho,Ohashi}, indicating that the system gradually becomes dominated by two-body bound molecules, as one passes through $(k_{\rm F}^3v)^{-1}= 0$. 
\par
\begin{figure}
\centerline{\includegraphics[width=7cm]{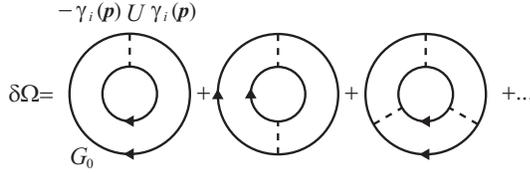}}
\caption{$p$-wave fluctuation corrections $\delta\Omega$ to the thermodynamic potential $\Omega$. The solid line and the dashed line represent, respectively, the bare Green' function $G_0({\bm p},i\omega_n)=(i\omega_n-\xi_{\bm p})^{-1}$, and a $p$-wave pairing interaction $V_p({\bm p},{\bm p}')$, where $\omega_n$ is the fermion Matsubara frequency.}
\label{fig1}
\end{figure}
\par
We include $p$-wave pairing fluctuations within the framework of the strong-coupling theory developed by Nozi\`eres and Schmitt-Rink (NSR). The advantage of this approach is that it directly evaluates fluctuation corrections $\delta\Omega$ to the thermodynamic potential $\Omega=\Omega_0+\delta\Omega$, that are diagrammatically given in Fig. \ref{fig1} (where $\Omega_{0}=T \sum_{{\bm p}} \ln [1+e^{-\xi_{\bm p}/T}]$ is the thermodynamic potential of a free Fermi gas). Once $\Omega$ is obtained, one can calculate the specific heat $C_V=(\partial E/\partial T)_{V,N}$ at constant volume, by evaluating the internal energy $E$ from the Legendre transformation,
\begin{equation}
E=
\Omega
-T\left( \frac{\partial \Omega }{\partial T} \right)_{\mu}
-\mu\left( \frac{\partial \Omega }{\partial \mu} \right)_{T}.
\label{eq.5}
\end{equation}
\par
Summing up the diagrams in Fig. \ref{fig1}, we obtain\cite{Inotani,Inotani2},
\begin{equation}
\delta\Omega=T\sum_{i\nu_n,{\bm q}}
{\rm Tr}\left[
\ln \hat{\Gamma}({\bm q},i\nu_n)
\right],
\label{eq.6}
\end{equation}
where $\nu_n$ is the boson Matsubara frequency, and 
\begin{equation}
{\hat \Gamma}({\bm q},i\nu_n)=-
{U \over 1-U{\hat \Pi}({\bm q},i\nu_n)}
\label{eq.7}
\end{equation}
is the $3\times 3$-matrix particle-particle scattering matrix, describing fluctuations in the $p$-wave Cooper channel. In Eq. (\ref{eq.7}), ${\hat \Pi}=\{\Pi_{i,j}\}$ ($i,j=x,y,z$) is the lowest-order pair correlation function, given by
\begin{equation}
\Pi_{i,j}({\bm q},i\nu_n)
=
-\frac{1}{2}\sum_{\bm {p}} 
\gamma^i_{\bm{p}}
\gamma^j_{\bm{p}} 
\frac{1-f(\xi_{\bm{p}+\frac{\bm q}{2}})-f(\xi_{\bm{p}-\frac{\bm q}{2}})}
{i\nu_n - \xi_{\bm{p} + \frac{\bm q}{2}} - \xi_{-\bm{p} + \frac{\bm q}{2}}}.
\label{eq.8}
\end{equation}
Here, $f(x)$ is the Fermi distribution function. We briefly note that by taking $z$ axis along the direction of ${\bm q}$, $\hat\Pi$, as well as $\hat{\Gamma}$ become diagonal in the present isotropic $p$-wave interaction above $T_{\rm c}$.
\par
To calculate $C_V$ in the normal state, we need to evaluate the Fermi chemical potential $\mu(T\ge T_{\rm c})$, which is, as usual, determined from the equation for the number $N$ of Fermi atoms\cite{NSR}, 
\begin{equation}
N=
-\left(
{\partial \Omega \over \partial \mu}
\right)_T=
-\left(
{\partial \Omega_0 \over \partial \mu}
\right)_T
-
\left(
{\partial \delta\Omega \over \partial \mu}
\right)_T.
\label{eq.8b}
\end{equation}
\par
We briefly note that, in the NSR scheme, the superfluid phase transition temperature $T_{\rm c}$ is determined from the Thouless criterion, $\Gamma^{-1}(0,0)=0$\cite{NSR,Inotani,Inotani2}.
\par
\begin{figure}
\centerline{\includegraphics[width=7cm]{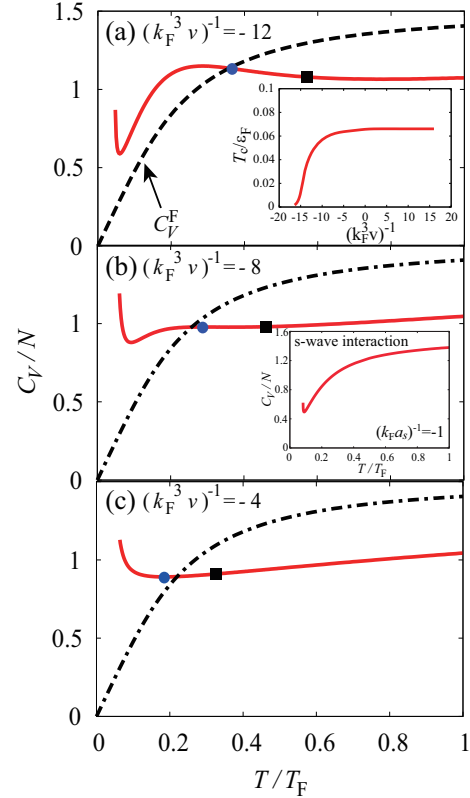}}
\caption{(Color online) Calculated specific heat $C_V$ at constant volume in the weak-coupling side ($(k_{\rm F}^3v)^{-1}\lesssim 0$) of a one-component $p$-wave interacting Fermi gas above $T_{\rm c}$. $C_V^{\rm F}$ is the specific heat in a free Fermi gas. The solid square and solid circle show the temperature at which $\mu(T)=0$ and $\mu(T)=T$, respectively (see also Fig. \ref{fig4}). The inset in panel (a) shows $T_{\rm c}$, as a function of the $p$-wave interaction strength. $\varepsilon_{\rm F}$ is the Fermi energy. The inset in panel (b) shows $C_V$ in the weak-coupling regime of an $s$-wave interacting Fermi gas, where $a_s$ is the $s$-wave scattering length.
}
\label{fig2}
\end{figure}
\par
Figure \ref{fig2} shows the specific heat $C_V$ at constant volume above $T_{\rm c}$ in the weak-coupling side of a $p$-wave interacting Fermi gas ($(k_{\rm F}^3v)^{-1}<0$, see also the inset in Fig. \ref{fig2}(a)). In panel (a), one clearly sees a dip-hump structure. Figures \ref{fig2}(b) and (c) show that this structure gradually becomes obscure, with increasing the interaction strength. In Fig. \ref{fig2}(c), while the dip still remains, the hump no longer exists. In a free Fermi gas, the specific heat $C_V$ exhibits a monotonic temperature dependence (dashed line in Fig. \ref{fig2}). In addition, such a hump structure is also absent in an $s$-wave interacting Fermi gas, as shown in the inset in Fig. \ref{fig2}(b). Thus, the $p$-wave interaction is found to play a crucial role for the appearance of the hump structure of $C_V$.
\par
\begin{figure}
\centerline{\includegraphics[width=7cm]{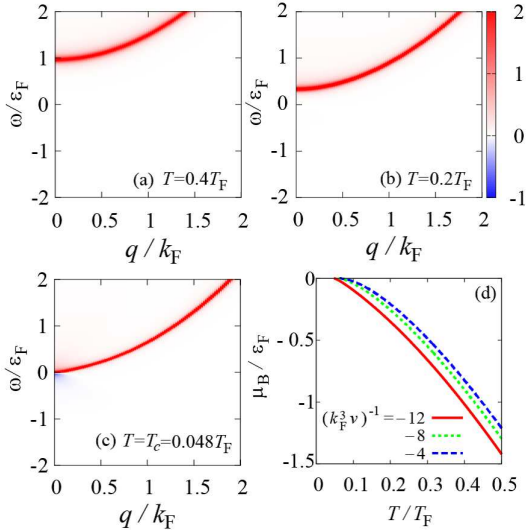}}
\caption{(Color online) (a)-(c) Calculated spectra of the analytic continued particle-particle scattering matrix, $-{\rm Im}[\Gamma({\bm q},i\nu_n \to \omega+i\delta)]$. The intensity is normalized by $\varepsilon_{\rm F}/k_{\rm F}^5$. The same normalization is also used in Figs. \ref{fig6}(a)-(c). We set $(k_{\rm F}^3 v)^{-1}=-12$. (d) Bose chemical potential $\mu_{\rm B}$, as a function of temperature.}
\label{fig3}
\end{figure}
\par
To understand the behavior of $C_V$ shown in Fig. \ref{fig2}, it is convenient to consider the single-particle thermal Green's function $G$ which is consistent with the NSR theory (in the sense that it gives the same number equation as Eq. (\ref{eq.8b})), given by
\begin{eqnarray}
G({\bm p},i\omega_n)
&=&
G_0({\bm p},\omega_n)
+G_0({\bm p},\omega_n)\Sigma({\bm p},\omega_n)G_0({\bm p},\omega_n)
\nonumber
\\
&\simeq&
{1 \over i\omega_n-\xi_{\bm p}-\Sigma({\bm p},i\omega_n)},
\label{eq.9}
\end{eqnarray}
where $G_0({\bm p},i\omega_n)=[i\omega_n-\xi_{\bm p}]^{-1}$ is the bare Green's function, $\omega_n$ is the fermion Matsubara frequency, and the self-energy $\Sigma({\bm p},i\omega_n)$ has the form\cite{Inotani},
\begin{equation}
\Sigma({\bm p},i\omega)=T
\sum_{{\bm q},i\nu_n}\sum_{i=x,y,z}
\gamma^i_{{\bm p}-{\bm q}/2}
\Gamma({\bm q},i\nu_n)
\gamma^i_{{\bm p}-{\bm q}/2}
G_0({\bm q}-{\bm p},i\nu_n-i\omega_n).
\label{eq.10}
\end{equation}
Near $T_{\rm c}$, since the particle-particle scattering matrix $\Gamma({\bm q},i\nu_n)$ is enhanced around ${\bm q}=\nu_n=0$\cite{NSR}, the self-energy in Eq. (\ref{eq.10}) may be approximated to\cite{Inotani}
\begin{equation}
\Sigma({\bm p},i\omega_n)\simeq -\Delta_{\rm {pg}}^2({\bm p})G_0(-{\bm p},-i\omega_n),
\label{eq.11}
\end{equation}
where $\Delta_{\rm pg}^2=-T{\bm p}^2\sum_{{\bm q},\nu_n}\Gamma({\bm q},i\nu_n)~(\ge 0)$ is the so-called pseudogap parameter\cite{Levin}, describing a particle-hole coupling by pairing fluctuations. When one substitutes Eq. (\ref{eq.11}) into the second line in Eq. (\ref{eq.9}), the resulting Green's function has the same form as the diagonal component of the mean-field BCS Green's function in the superfluid phase as,
\begin{equation}
G({\bm p},i\omega_n)=-
{i\omega_n+\xi_{\bm p} 
\over \omega_n^2+\xi_{\bm p}^2+\Delta_{\rm pg}^2({\bm p})
}.
\label{eq.12}
\end{equation}
Equation (\ref{eq.12}) indicates that the system near $T_{\rm c}$ has superfluid-phase-like properties, which is sometimes referred to as the pseudogap phenomenon in the literature. As a result, the entropy $S$ is suppressed near $T_{\rm c}$, leading to the enhancement of $C_V=T(\partial S/\partial T)_{V,N}$ as seen in Fig. \ref{fig2} below the dip structure.
\par
\begin{figure}
\centerline{\includegraphics[width=7cm]{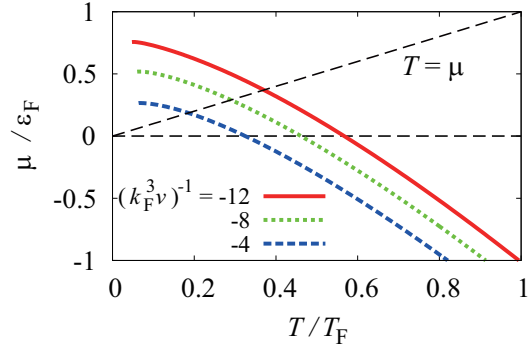}}
\caption{(Color online) Calculated Fermi chemical potential $\mu$, as a function of temperature. We note that, in each panel in Fig. \ref{fig2}, the solid square and solid circles are, respectively, the temperature at which $\mu(T)=0$ and the temperature at which $\mu(T)$ intersects the dashed line ($\mu=T$).}
\label{fig4}
\end{figure}
\par
We note that the same pseudogap phenomenon is also seen in the $s$-wave case (inset in Fig. \ref{fig2}(b))\cite{Pieter}. Thus, the enhancement of $C_V$ near $T_{\rm c}$ and the resulting dip structure seen in Fig. \ref{fig2} should be regarded as a phenomenon widely in attractively interacting Fermi gases near $T_{\rm c}$, rather than a phenomenon peculiar to the $p$-wave case.
\par
On the other hand, the key to understand the hump structure seen in Fig. \ref{fig2}(a) is that the relatively large magnitude of the inverse effective range $|k_0/k_{\rm F}|=30\gg 1$\cite{Ticknor}, as well as the momentum dependence of the $p$-wave interaction ($V_p({\bm p},{\bm p}')\propto pp'$), lead to the Bose-propagator-like structure of the particle-particle scattering matrix $\Gamma({\bm q},i\nu_n)$ as\cite{Inotani}
\begin{equation}
\Gamma({\bm q}, i\nu_n)\simeq
{Z \over i\nu_n-E_{\bm q}},
\label{eq.13}
\end{equation}
where $Z=24\pi/(m^2|k_0|)$, and $E_{\bm q}={\bm q}^2/(4m)-\mu_{\rm B}$ with $\mu_{\rm B}=2\mu-2/(mvk_0)$. Indeed, as shown in Figs. \ref{fig3}(a)-(c), the spectrum $-{\rm Im}[\Gamma({\bm q},i\nu_n\to \omega+i\delta)]$ has a sharp peak line along the molecular dispersion $\omega=E_{\bm q}$ (where $\delta$ is an infinitesimally small positive number). Substituting Eq. (\ref{eq.13}) into Eq. (\ref{eq.11}), the Fermi quasiparticle damping rate, ${\bar \gamma}({\bm p},\omega)=-{\rm Im}[\Sigma({\bm p},i\omega_n\to\omega+i\delta)]$, is evaluated as
\begin{eqnarray}
{\bar \gamma}({\bm p},\omega)
&=&
-\pi Z
\sum_{\bm q}
\left[
{\bm p}+{{\bm q} \over 2}
\right]^2
\nonumber
\\
&\times&
\left[
f(\xi_{{\bm q}-{\bm p}})+
n_{\rm B}(E_{\bm q})
\right]
\delta(\omega-E_{\bm q}+\xi_{{\bm q}-{\bm p}}).
\label{eq.14}
\end{eqnarray}
Here, $n_{\rm B}(x)$ is the Bose distribution function. For simplicity, when we consider ${\bar \gamma}({\bm q},\omega=0)$ in the case of Fig. \ref{fig2}(a), Eq. (\ref{eq.14}) indicates that, with decreasing the temperature from the high temperature region where $\mu<0$, the Fermi quasiparticle scattering starts to be suppressed, when $\mu$ becomes positive and most Fermi atoms are occupied below $\mu$. This is because the constraint $\xi_{{\bm q}-{\bm p}}=E_{\bm q}\ge |\mu_{\rm B}|\sim\varepsilon_{\rm F}$ (see Fig. \ref{fig3}(d)) coming from the $\delta$-function in Eq. (\ref{eq.14}) becomes difficult to be satisfied. This suppression decreases the entropy $S$, which naturally leads to the increase of the specific heat $C_V=T(\partial S/\partial T)_{V,N}$ with decreasing the temperature. Indeed, Fig. \ref{fig2}(a) shows that the hump structure of $C_V$ starts to appear around the temperature below which the chemical potential $\mu$ becomes positive (solid square). In addition, once the Fermi degeneracy is achieved, the temperature dependence of the suppression of quasiparticle scattering, as well as the temperature dependence of the suppression of the entropy $S$, would become weak. At such low temperatures, $C_V$ is expected to decrease with decreasing the temperature, as in the case of a degenerate free Fermi gas (except near $T_{\rm c}$). Indeed, the temperature at the top of the hump structure in Fig. \ref{fig2}(a) is close to the temperature which satisfies $T=\mu(T)$ (solid circle).   
\par
\begin{figure}
\centerline{\includegraphics[width=7cm]{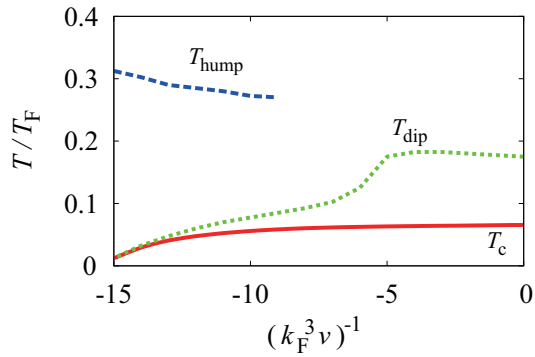}}
\caption{Two characteristic temperatures $T_{\rm hump}$ and $T_{\rm dip}$ that are determined as the temperature at the top of the hump structure and the temperature at the dip of the specific heat $C_V$, respectively. $T_{\rm c}$ is the $p$-wave superfluid phase transition temperature.}
\label{fig5}
\end{figure}

With increasing the interaction strength, the pseudogap regime where $C_V$ is enhanced becomes wide. At the same time, the temperature below which the above-mentioned mechanism of the hump structure works becomes low, because the temperature at which $\mu$ changes its sign becomes low, as shown in Fig. \ref{fig4}. These naturally explain why the hump structure of $C_V$ gradually disappears, as one approaches the intermediate coupling regime from the weak-coupling side. When we plot the temperature $T_{\rm hump}$ at the top of the hump structure, we numerically confirm that $T_{\rm hump}$ disappears at $(k_{\rm F}^3v)^{-1}\simeq -8$, as shown in Fig. \ref{fig5}. On the other hand, when we plot the temperature $T_{\rm dip}$ at the dip, we find in Fig. \ref{fig5} that it continues to exist even in the intermediate coupling regime.
\par
\par
\begin{figure}
\centerline{\includegraphics[width=7cm]{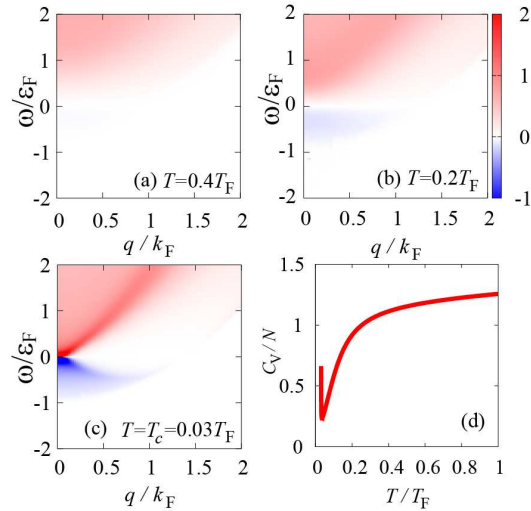}}
\caption{(Color online) (a)-(c) Calculated spectrum of the analytic continued particle-particle scattering matrix, $-{\rm Im}[\Gamma({\bm q},i\nu_n \to \omega+i\delta)]$, when the basis function $\gamma^i_{\bm p}=({\bm p}/|{\bm p}|)F_{\rm c}({\bm p})$ is used. We take $(k_{\rm F}^3 v)^{-1}=-1$. The panel (d) shows $C_V$.}
\label{fig6}
\end{figure}

We note that all the $p$-wave pairing interactions do not give the dip-hump behavior of $C_V$. Indeed, when we only retain the $p$-wave symmetry but ignore the $|{\bm p}|$-dependence of the prefactor of $F_{\rm c}({\bm p})$ in the basis function as $\gamma^i_{\bm p}=({\bm p}/|{\bm p}|)F_{\rm c}({\bm p})$, $C_V$ does not exhibit the hump structure, as shown in Fig. \ref{fig6}(d). In the case of $\gamma^i_{\bm p}={\bm p}F_{\rm c}({\bm p})$, the $p$-wave interaction is enhanced in the high-energy region by the factor ${\bm p}\cdot{\bm p}'$ in $V_p({\bm p},{\bm p}')$ in Eq. (\ref{eq.1b}), which also enhances the molecular character of pairing fluctuations, leading to the sharp peak line along $E_{\bm q}$ in Figs. \ref{fig3}(a)-(c). In contrast, this effect is absent in the case of $\gamma^i_{\bm p}=({\bm p}/|{\bm p}|)F_{\rm c}({\bm p})$. Thus, the spectrum of the particle-particle scattering matrix $\Gamma({\bm q},i\nu_n\to\omega+i\delta)$ spreads out, as shown in Figs. \ref{fig6}(a)-(c). In this case, the constraint coming from the $\delta$-function in Eq. (\ref{eq.14}) is not obtained. As a result, even when the Fermi chemical potential $\mu$ becomes positive and the Fermi degeneracy starts to occur, the suppression of the entropy $S$ is not remarkable enough to cause the enhancement of $C_V$.
\par
Since the ordinary contact-type $s$-wave pairing interaction also does not have a factor which enhances the interaction strength in the high-energy region, the situation is similar to the case of $\gamma^i_{\bm p}=({\bm p}/|{\bm p}|)F_{\rm c}({\bm p})$. Thus, the spectrum of $\Gamma({\bm q},i\nu_n\to\omega+i\delta)$ also spreads out in this case, although we do not explicitly show the result here. This explains the reason for the absence of the hump structure in the $s$-wave case shown in the inset in Fig. \ref{fig2}(b).
\par
\par
Strictly speaking, the above-mentioned mechanism for the hump structure would also work for other unconventional pairing interactions, such as a $d$-wave one. Thus, the observation of a dip-hump structure of $C_V$ does not immediately mean that a $p$-wave interaction play an essential role. However, the advantage of an ultracold Fermi gas is that one can unambiguously pick up the symmetry of pairing interaction by using a Feshbach resonance. Thus, when a hump structure of $C_V$ is observed near a $p$-wave Feshbach resonance, we may actually conclude that it comes from the $|{\bm p}|$-dependence of a $p$-wave pairing interaction associated with the $p$-wave Feshbach resonance. 
\par
To summarize, we have theoretically discussed the specific heat $C_V$ at constant volume in the weak-coupling side of a $p$-wave interacting Fermi gas. Within the framework of the strong-coupling theory developed by Nozir\`eres and Schmitt-Rink, we showed that a $p$-wave interaction causes a dip-hump behavior of $C_V$ as a function of the temperature above $T_{\rm c}$. While the dip structure, which is related to the pseudogap phenomenon, also appears in an $s$-wave interacting Fermi gas, the hump structure does not. Thus, the latter may be viewed as a characteristic phenomenon in a $p$-wave interacting Fermi gas. Since the achievement of the $p$-wave superfluid phase, which is the most characteristic phenomenon in a $p$-wave interacting Fermi gas, seems still difficult in the current stage of cold Fermi gas physics, the observation of this normal-state phenomenon expected in the weak-coupling side of a $p$-wave interacting Fermi gas would be helpful to clarify how this system is qualitatively different from the ordinary $s$-wave system that has extensively been studied in this field.
\par
\begin{acknowledgment}
This work was supported by KiPAS project in Keio University, as well as Grant-in-aid for Scientific Research from MEXT and JSPS in Japan (No.JP16K17773, No.JP15H00840, No.JP15K00178, No.JP16K05503). 
\end{acknowledgment}


\end{document}